\documentclass[%
reprint,
superscriptaddress,
amsmath,
amssymb,
aps,
prl,
floatfix,
]{revtex4-2}

\usepackage{float}
\usepackage{xcolor}
\usepackage{graphicx}
\usepackage{dcolumn}
\usepackage{bm}
\usepackage{comment}
\usepackage{soul}
\usepackage{siunitx}

\usepackage[unicode=true,pdfusetitle,bookmarks=false,colorlinks=true,citecolor=blue,urlcolor=blue,linkcolor=blue]{hyperref}

\begin{document}

\preprint{APS/123-QED}

\title{Single-File Diffusion of Active Brownian Particles}

\author{Akinlade Akintunde}
\affiliation{Department of Chemistry, The Pennsylvania State University, University Park, Pennsylvania, 16802, USA}

\author{Parvin Bayati}
\affiliation{Department of Chemistry, The Pennsylvania State University, University Park, Pennsylvania, 16802, USA}

\author{Hyeongjoo Row}
\affiliation{Department of Chemical and Biomolecular Engineering, UC Berkeley, Berkeley, CA 94720, USA}

\author{Stewart A. Mallory}
\email{sam7808@psu.edu}
\affiliation{Department of Chemistry, The Pennsylvania State University, University Park, Pennsylvania, 16802, USA}
\affiliation{Department of Chemical Engineering, The Pennsylvania State University, University Park, Pennsylvania, 16802, USA}


\date{\today}

\begin{abstract}
Single-file diffusion (SFD) is a key mechanism underlying transport phenomena in confined physical and biological systems.  
In a typical SFD process, microscopic particles are restricted to moving in a narrow channel where they cannot pass one another, resulting in constrained motion and anomalous long-time diffusion.  
In this study, we use Brownian dynamics simulations and analytical theory to investigate the SFD of {\color{black} athermal} active Brownian particles (ABPs)—a minimal model of active colloids.  
{\color{black} Building on prior work [\textit{Phys. Rev. E} \textbf{108}, 064601 (2023)], where the kinetic temperature, pressure, and compressibility of the single-file ABP system were derived}, we develop an accurate analytical expression for the mean square displacement (MSD) of a tagged particle.  
We find that the MSD exhibits ballistic behavior at short times, {\color{black} governed by the reduced kinetic temperature of the system}.  
At long times, the characteristic subdiffusive scaling of SFD, $[\langle (\Delta x)^2 \rangle \sim t^{1/2}]$, is preserved.  
However, self-propulsion introduces significant changes to the 1D-mobility, {\color{black} which we directly relate to the system's compressibility}.  
Furthermore, we demonstrate that the generalized 1D-mobility, originally proposed by Kollmann for equilibrium systems [\textit{Phys. Rev. Lett.} \textbf{90}, 180602 (2003)], can be extended to active systems with minimal modification.  
These findings provide a framework for understanding particle transport in active systems and for tuning transport properties at the microscale, particularly in geometries where motion is highly restricted.
\end{abstract}

\maketitle

\section{Introduction}

\begin{figure}[t!]
	\centering
	\includegraphics[width=.475\textwidth]{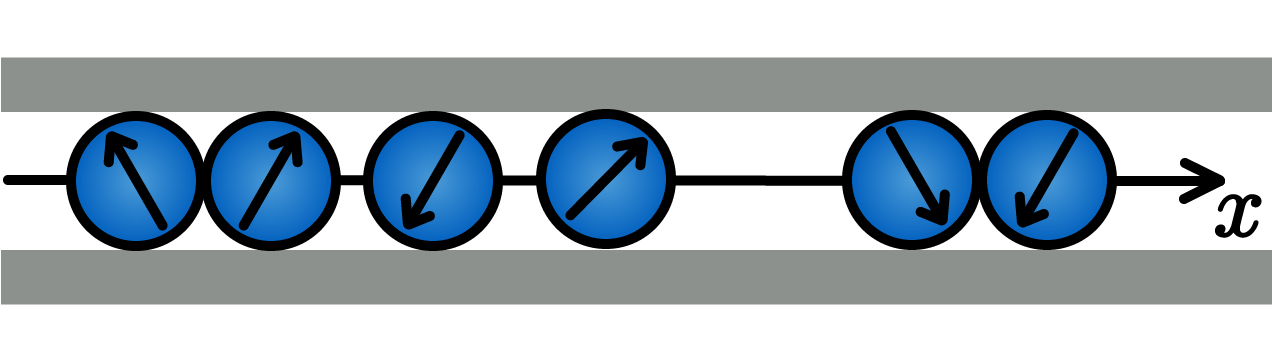}
	\caption{\protect\small{Schematic of SF-ABP system. Each purely repulsive active particle moves at a constant speed of $U_a \cos \theta$ while undergoing rotational Brownian motion with a reorientation time of $\tau_R$.}
    }
	\label{figure_1}
\end{figure}

For over half a century, single-file diffusion (SFD) has captivated researchers across the physical and biological sciences, finding broad applications in quantifying transport at the microscale~\cite{Hodgkin1955-kb, Jepsen1965-qg, Lebowitz1967-ml, Levitt1973-hj, Stephan1983-ti, Eskesen1986-lo, MacElroy1997-gk, Rodenbeck1998-mc, Mon2002-au, Kollmann2003-as, Nelissen2007-sd, Lizana2008-pz, Kharkyanen2009-zp, Herrera-Velarde2010-le, Stern2013-qp, Leibovich2013-my, Goldt2014-tj, Locatelli2016-nw, Farrell2021-ig}.  
In a typical SFD process, microscopic particles are confined to move within a narrow channel where they cannot pass one another (See Fig.~\ref{figure_1} for a schematic illustration of SFD).  
This form of confinement leads to subdiffusive behavior at long times, making SFD one of the simplest examples of anomalous diffusion~\cite{Harris1965-cb, Levitt1973-hj, Kollmann2003-as, Karger2008-kg, Felderhof2009-ag, Centres2010-qb, Hofling2013-ok, Jobic2016-ow}.  
As a result, SFD provides a fundamental framework for understanding transport in highly confined systems, with far-reaching implications for both theoretical studies and practical applications.

Most studies of SFD have focused on equilibrium Brownian particles (i.e., passive particles), where it is nearly universally observed that, at long times, the mean square displacement (MSD) of a tagged particle scales as
\begin{equation}
    \label{eq:1}
    \lim_{t \rightarrow \infty} \langle (\Delta x)^{2} \rangle = 2 F t^{1/2} \, ,
\end{equation}
where $F$ is a proportionality constant known as the 1D-mobility.  
This characteristic subdiffusive scaling of SFD has been thoroughly explored computationally and experimentally in various systems, including colloidal particles in microchannels~\cite{Wei2000-si, Lutz2004-gy, Lutz2004-lo, Lin2005-uh, Coste2010-xa, Villada-Balbuena2021-po} and molecular transport in zeolites and carbon nanotubes~\cite{Kukla1996-ct, Karger1997-cn, Karger2008-ta, Li2023-ev}.  
In addition to experimental studies, several theoretical approaches have been developed to describe the SFD of passive Brownian systems, including fluctuating hydrodynamics~\cite{Boldrighini1983-gi, Bleibel2014-kj, Rizkallah2023-om}, fractional Langevin formalisms~\cite{Taloni2008-rx, EAB20102510, Taloni2014-pg}, and asymptotic methods~\cite{temporaryCitekey-ahhcb, Hahn1995-ct, Flomenbom2008-wb}.  
We refer the reader to Refs~\cite{Centres2010-qb, Lomholt2014-we, temporaryCitekey-iebdu, Benichou2018-ad}, for a more comprehensive discussion on passive SFD.

The central theoretical challenge in SFD is to capture the many-body effects that give rise to subdiffusive behavior at long times and to quantitatively predict the 1D-mobility.  
In one of the earliest studies of SFD, Harris~\cite{Harris1965-cb} derived the fundamental result for the 1D-mobility of passive hard particles
\begin{equation}
    \label{eq:2}
    F_{HR} = \frac{1-\phi}{\rho} \sqrt{\frac{D}{\pi}} = \lambda \sqrt{\frac{D}{\pi}} \, ,
\end{equation}
where $D$ is the free diffusion coefficient of the particle, $\phi = \rho \sigma_p$ is the packing fraction, $\sigma_p$ is the particle diameter, and $\rho$ is the particle line density.  
The mean free path $\lambda$, which represents the average distance a particle can move before encountering its neighbor, can be calculated analytically for single-file passive hard particles and is given by $\lambda = (1 - \phi)/\rho$~\cite{Santos2016-gn}.  
In Harris's derivation of the 1D-mobility, the mean free path is a fundamental quantity that provides a measure of the confinement experienced by a tagged particle and can be introduced to further simplify Eq.~(\ref{eq:2}). 
We also find the mean free path plays a similar pivotal role in the SFD of active colloids. 

More recently, using an asymptotic approach, Kollmann generalized the 1D-mobility for passive particles to systems with arbitrary finite-range interactions~\cite{Kollmann2003-as}.  
In such systems, the 1D-mobility is expressed as
\begin{equation}
    \label{eq:3}
    F = \frac{1}{\rho} \sqrt{\frac{D \mathcal{X}}{\pi}} \, ,
\end{equation}
where $\mathcal{X}$ is the reduced isothermal compressibility, defined as the ratio between the compressibility of an interacting system to that of an ideal system.  
We should note that for more complicated interparticle interactions, the quantity $\sqrt{\mathcal{X}}/ \rho$ plays a similar role to the mean free path introduced in Eq.~(\ref{eq:2}).  
This generalized result has been verified across various systems, including attractive particles, mixtures, and particles with long-ranged interactions~\cite{Lutz2004-gy, Coste2010-xa, Lizana2010-ag}.  
Interestingly, the work of Kollmann and others illustrates a peculiar feature of single-file systems: the 1D-mobility, a single-particle transport property, can be related to a thermodynamic response function of the system (i.e., the isothermal compressibility). 

Much of the prior work on SFD has focused on passive particles, motivating the natural question: how is single-file diffusion altered when particle dynamics deviate from passive Brownian motion~\cite{Harris1965-cb, Jepsen1965-qg, Levitt1973-hj, Kollmann2003-as}? 
To answer this question, we focus on active colloids capable of self-propulsion at the microscale. 
Interestingly, this novel class of colloids exhibits persistent directed motion at short times and undergoes diffusive motion at long times.  
The self-driven nature of active colloids has spurred extensive investigations into their phenomenology and use in potential applications ranging from transport, self-assembly, and fluid mixing at the microscale~\cite{Redner2013-nn, Fily2012-ej, Illien2013-tv, Soto2014-if, Mallory2014-qn, Mallory2015-fy, Mallory2016-mc, Mallory2017-yv, Mallory2018-nn, Barberis2019-le, Mallory2019-to, Mallory2020-cx, Mallory2021-ux, Omar2023-bq, Paul2024-wx, Bayati2024-km}.  
This includes several studies that have explored different aspects of single-file active matter systems~\cite{Romanczuk2010-hi, Locatelli2014-is, Fily2014-do, Euan-Diaz2014-nc, temporaryCitekey-pfkun, Misiunas2019-pm, Dolai2020-cu, Banerjee2022-zb, Jose2022-fn, Caprini2020-gj, Marconi2024-gk}.
However, there has been no systematic attempt to quantify the role of activity in single-file systems and how it influences observables such as the 1D-mobility. 
Understanding the SFD of active colloids is critical, as many proposed applications operate in environments where such conditions are likely.

In this work, we investigate the single-file diffusion (SFD) of active colloids within the context of the {\color{black} athermal} active Brownian particle (ABP) model.  
Here, we analyze the long- and short-time behavior of the mean square displacement (MSD), focusing on how self-propulsion modifies both the ballistic motion at short times and the subdiffusive scaling at long times.  
Our results demonstrate that Kollmann’s generalized 1D-mobility expression for passive systems [Eq.~(\ref{eq:3})] can be extended to active systems with minimal modification.  
This work provides new insights into the behavior of confined active matter informing the design of microfluidic systems and targeted transport technologies, where precise control of particle dynamics is essential. 

\subsection{Model}

In this study, we consider a periodic single-file active Brownian particle (SF-ABP) model, {\color{black}which was introduced in our prior work}~\cite{Schiltz-Rouse2023-vj}, where $N$ purely-repulsive active Brownian disks are confined to move along a narrow channel of length $L$, as depicted in Fig.~\ref{figure_1}.  
The channel is sufficiently narrow to prevent particles from passing one another, enforcing a single-file condition that restricts their motion to one spatial dimension.  
Each particle experiences a self-propelling force $F_a = \gamma U_a \cos{\theta}$, where $\theta$ is the angle between the particle's orientation vector and the positive x-axis, $\gamma$ is the translational drag coefficient, and $U_a$ is the constant self-propelling speed.  
The orientation of each particle undergoes rotational Brownian motion with a characteristic reorientation time $\tau_R$.  
The following overdamped equations describe the motion of a tagged particle:
\begin{subequations}
    \label{eq:4}
    \begin{equation}
        v = \dot{x} = U_a \cos\theta + \gamma^{-1} F_c \ ,
    \end{equation}
    \begin{equation}
        \dot{\theta} = \xi(t) \ ,
    \end{equation}
\end{subequations}
\noindent where $F_c$ represents the interparticle forces, and $\xi(t)$ is the stochastic rotational noise with properties $\langle \xi(t) \rangle = 0$ and ${\langle \xi(t)\xi(t') \rangle = (2/\tau_R)\delta(t-t')}$.
For simplicity, we assume translational Brownian motion {\color{black} (i.e., athermal)} and hydrodynamic interactions are negligible. 
{\color{black} The effects of translational Brownian motion can be incorporated in a straightforward manner by including the standard Gaussian white noise to [Eq.~(\ref{eq:4}a)].
}

The interparticle forces $F_c$ are implemented using a Weeks-Chandler-Anderson (WCA) potential, characterized by a potential strength $\varepsilon$ and Lennard-Jones diameter $\sigma$~\cite{Weeks1971-wy}.  
We choose the potential strength to be sufficiently large (i.e., $\varepsilon/(F_a \sigma) = 100$) to effectively mimic hard-particle behavior endowing particles with an effective hard diameter of $\sigma_p = 2^{1/6} \sigma$.  
For the initial configuration, particles are randomly placed in the channel without overlaps, and the initial angles for the orientation vectors are uniformly distributed.  
Simulations were performed using \texttt{HOOMD-blue}~\cite{Anderson2020-qc} with N = 1000 particles for a minimum of $\num{1.5e10}$ timesteps, and a timestep size of $\delta t = 10^{-5}$ time units.

\begin{figure*}[t!]
	\includegraphics[width=0.95\textwidth]{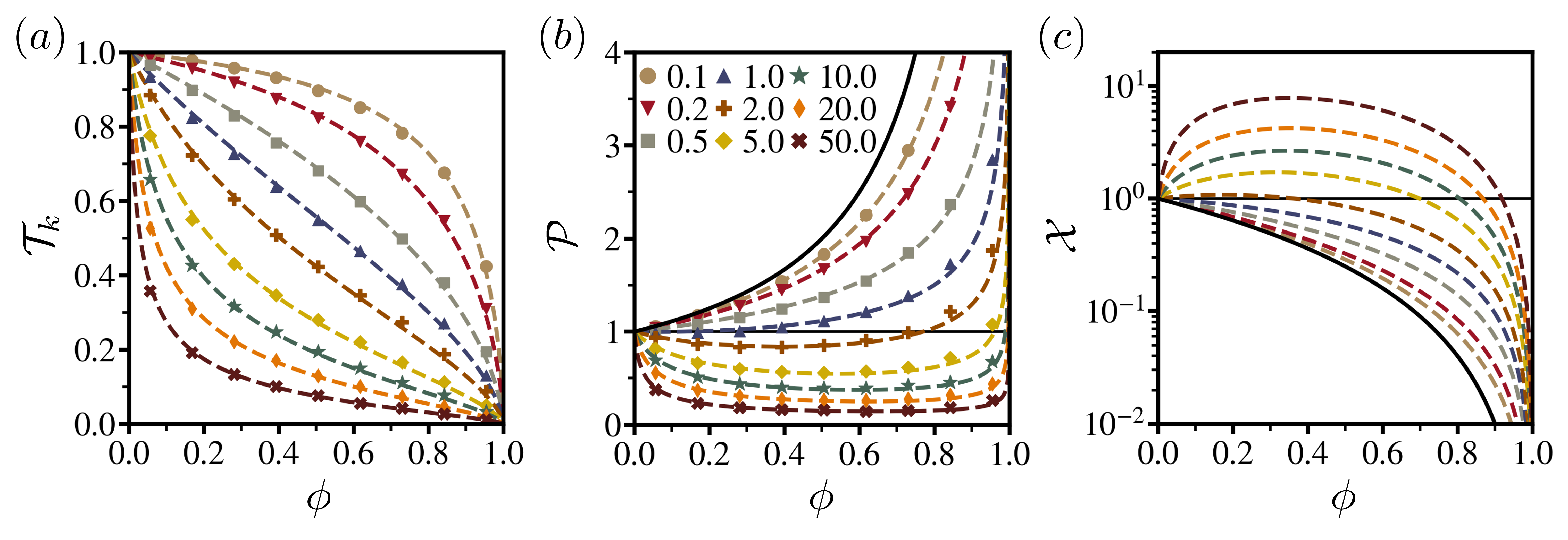}        
	\caption{\protect\small{{\color{black} Brownian dynamics simulation results (points) for the reduced (a) kinetic temperature and (b) pressure at different active Péclet numbers vs packing fraction. In panel (c), the constant Pe compressibility is only computed from the analytical expression for the pressure. However, the compressibility can be computed directly from simulation via the static structure factor (See Ref.~\cite{Dulaney2021-kf} for details). The dashed colored lines in (a)-(c) represent the analytical expressions for these quantities [Eqs.~(\ref{eq:6})-(\ref{eq:8})]. The solid line in panels (b) and (c) corresponds to the analytical expression for the reduced pressure and isothermal compressibility of the 1D passive hard particle system (Tonks gas).
}}}
	\label{figure_2}
\end{figure*}

The state of the SF-ABP system is characterized by two dimensionless parameters: the packing fraction $\phi = \rho \sigma_p$, where $\rho = N/L$ is the particle line density, and the active P\'eclet number $\text{Pe} = \ell_0/\sigma$, where $\ell_0 = U_a \tau_R$ is the intrinsic run length of a particle.  
The active P\'eclet number quantifies the persistence of a particle's motion relative to its diameter.  
{\color{black} We note that particle motion is entirely due to self-propulsion in the athermal ABP model, and therefore, we consider $\text{Pe}>0$. 
In the limit $\text{Pe} \ll 1$, particles undergo nearly Brownian dynamics, and we recover the mechanical, structural, and transport properties of the single-file hard particle system often referred to as the Tonks gas~\cite{Tonks1936-pf} with temperature $T\propto{\text{Pe}^{2}}$.
The case of $\text{Pe}=0$ corresponds to the singular limit of a classical system at $T=0$—a static system.}  
As $\text{Pe}$ increases, large dynamic clusters begin to form~\cite{Dolai2020-cu,Gutierrez2021-mu, Debnath2023-uy}.  
A detailed analysis of clustering behavior in the SF-ABP system is provided in Ref.~\cite{Sepulveda2016-sr, Dolai2020-cu, de-Castro2021-xt, Schiltz-Rouse2023-vj}.  
Interestingly, despite significant clustering, the SF-ABP system does not exhibit a motility-induced phase transition at finite P\'eclet numbers~\cite{Locatelli2015-jl, Dolai2020-cu, Gutierrez2021-mu, Mukherjee2023-ca}.

{\color{black} Here, we provide a brief summary of the concepts of kinetic temperature, pressure, and compressibility derived for the SF-ABP system, which are essential for characterizing the mean square displacement (MSD) of a tagged particle.
The full derivations of these quantities can be found in our prior work~\cite{Schiltz-Rouse2023-vj}.  
}
The reduced kinetic temperature $\mathcal{T}_k = 2 \langle v^2 \rangle / U_a^2$, which represents the ratio of the mean square velocity in the interacting system to that of an ideal ABP system, is expressed as:
\begin{equation}
    \label{eq:5}
    \mathcal{T}_k = 1 - \frac{2 \langle F_c^2\rangle}{(\gamma U_a)^2} \, ,
\end{equation}
where $\mathcal{T}_k$ quantifies the reduction in the mean square velocity due to interparticle collisions.
{\color{black} The concept of temperature in active systems has recently attracted significant interest, with a recent article discussing various definitions and their utility~\cite{Hecht2024-uz}.}  
In Fig.~\ref{figure_2}(a), we show simulation results (points) for the reduced kinetic temperature of the SF-ABP system.  
As $\phi \rightarrow 0$, $\mathcal{T}_k \rightarrow 1$, and as $\phi \rightarrow 1$, $\mathcal{T}_k \rightarrow 0$.  
Using scaling arguments based on collision timescales, we previously derived the following accurate expression for the reduced kinetic temperature [dashed lines in Fig.~\ref{figure_2}(a)]:
\begin{equation}
    \label{eq:6}
    \mathcal{T}_k = \frac{1}{9b^2}\left[2 \cos \left(\frac{1}{3}\arccos\left(\frac{27}{2} b^2 - 1\right)\right) - 1\right]^2 \ ,
\end{equation}
where $b = \alpha \text{Pe} \phi /(1-\phi)$ and $\alpha = c/(1+\text{Pe})^d$ with $c = 1.1$ and $d=0.05$.
Furthermore, we demonstrated that the reduced kinetic temperature is equivalent to the reduced swim pressure, $\mathcal{T}_k = P_s/P_0$, and derived an analytical expression for the reduced total pressure of the SF-ABP system:
\begin{equation}
    \label{eq:7}
    \mathcal{P} = \frac{P}{P_0} = \mathcal{T}_k\left[\frac{1}{1-\phi}\, \right] \ ,    
\end{equation}
where $P_0 = \rho \gamma U_a^2 \tau_R/2$ is the ideal gas pressure.  
In the limit of small $\text{Pe}$, $\mathcal{T}_k \rightarrow 1$ and Eq.~(\ref{eq:7}) reduces to the well-known equilibrium result for the Tonks gas~$\mathcal{P} = 1/(1-\phi)$~\cite{Tonks1936-pf}.  
Fig.~\ref{figure_2}(b) shows a comparison between the reduced pressure for the SF-ABP system and Eq.~(\ref{eq:7}).

The reduced constant P\'eclet compressibility, $\mathcal{X}$, [plotted in Fig.~\ref{figure_2}(c)] serves as a thermodynamic-like response function, analogous to the reduced isothermal compressibility in passive systems~\cite{Dulaney2021-kf, Schiltz-Rouse2023-vj}.
This response function measures clustering and local density fluctuations, and as we show, is an important quantity in characterizing the MSD of a tagged particle.
It can be calculated directly from Eq.~(\ref{eq:7}), yielding  
\begin{equation}
    \label{eq:8}
    \mathcal{X} = \frac{\chi_{a}}{\chi_{0}} = \left[\frac{\mathcal{T}_k}{(1-\phi)^2} + \frac{\phi}{1-\phi} \left( \frac{\partial \mathcal{T}_k}{\partial \phi} \right) \right]^{-1},
\end{equation}
where $\chi_a = (\partial \ln \rho / \partial P)_{\text{Pe}}$ and $\chi_0 = 1/P_0$ are the compressibilities of the interacting and ideal systems, respectively.  
As $\text{Pe} \ll 1$, the SF-ABP system recovers the passive result, $\mathcal{X} = (1-\phi)^2$, and as $\phi \rightarrow 0$, $\mathcal{X} \rightarrow 1$.  

\begin{figure*}[t!]
	\includegraphics[width=0.65\textwidth]{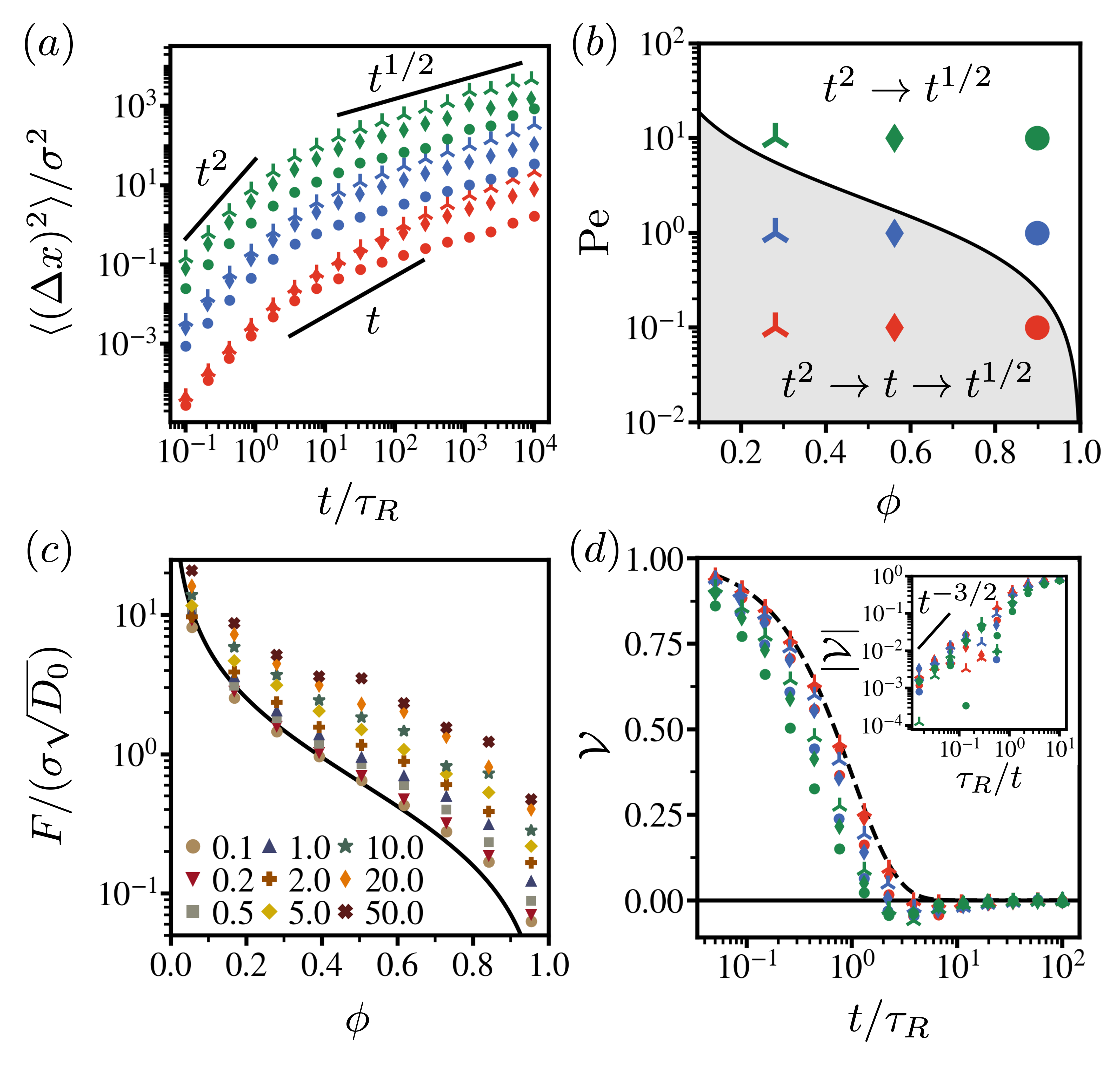}        
	\caption{\protect\small{(a) Mean square displacement (MSD) of a tagged particle in the SF-ABP system. (b) The state diagram for the SF-ABP system, {\color{black} with points corresponding to the curves shown in panels (a) and (d)}. The black line represents the condition $\ell_0/\lambda_{ABP} = 1$, {\color{black} which approximately separates two different dynamical regimes as discussed in the text}. (c) Simulation results for the 1D-mobility across different values of $\text{Pe}$ and $\phi$. {\color{black} The symbols and corresponding colors delineate different values of Pe}. The solid black line represents the analytical result for the 1D-mobility of passive hard particles [Eq.~(\ref{eq:2})]. (d) Normalized velocity autocorrelation function for the representative values of $\text{Pe}$ and $\phi$. The {\color{black} dashed} black line represents the analytical result for ideal {\color{black} 1D-ABPs}. The inset shows the power-law decay of the normalized velocity autocorrelation function, demonstrating the characteristic $-t^{-3/2}$ behavior.
}}
	\label{figure_3}
\end{figure*}
Before presenting the results of this study, we briefly review the mean square displacement (MSD) of an ideal suspension of {\color{black} 1D-ABPs}.  
A useful form of the MSD is
\begin{equation}
    \label{eq:9}
    \langle (\Delta x)^2 \rangle = 2 \langle v^2 \rangle t \left[ \int_0^t \left(1 - \frac{\tau}{t} \right) \mathcal{V}(\tau) d\tau \right] \,,
\end{equation}
where $\mathcal{V}(\tau) = \langle v(0)v(\tau) \rangle/\langle v^2 \rangle$ is the normalized velocity autocorrelation function and $\langle v^2 \rangle$ is the mean square velocity of the particle~\cite{Graham2018-ch}.  
In the ideal {\color{black}1D}-ABP system, where there are no interparticle forces ($F_c = 0$), it is straightforward to show that $\langle v^2 \rangle = \frac{1}{2}U_a^2$ and $\mathcal{V}(\tau) = e^{-\tau/\tau_R}$.  
Substituting these two expressions into Eq.~(\ref{eq:9}), we obtain the well-known result for ideal {\color{black} 1D-ABPs}~\cite{Hagen2009NonGaussianBO}:
\begin{equation}
    \label{eq:10}
    \langle (\Delta x)^2 \rangle = 2 D_0\left[t + \tau_{R}\left(e^{-t/\tau_{R}} - 1\right)\right],
\end{equation}
where $D_0 = \frac{1}{2} U_a^2 \tau_R$ is the free diffusion coefficient of an ABP.  
The MSD of an ideal ABP exhibits two distinct scaling regimes.  
At short times, the motion is ballistic, scaling as $\langle (\Delta x)^2 \rangle \sim \frac{1}{2} U_a^2 t^2$.  
At long times, the motion becomes diffusive, scaling as $\langle (\Delta x)^2 \rangle \sim 2 D_0 t$.  
The crossover between the ballistic and diffusive regimes occurs around $t \approx \tau_R$, corresponding to the timescale over which the particle's orientation undergoes significant reorientation due to rotational diffusion.

\section{Results}

Figure~\ref{figure_3} summarizes the simulation results for the SF-ABP system.  
The mean square displacement (MSD) for a representative selection of packing fractions and P\'eclet numbers is shown in Fig.~\ref{figure_3}(a).  
For all values of $\text{Pe}$ and $\phi$, the MSD exhibits ballistic behavior at times less than the reorientation time $\tau_R$. 
At sufficiently long times, for all values of $\text{Pe}$ and $\phi$, the MSD transitions to the characteristic subdiffusive scaling of single-file diffusion, where $\langle x^2(t) \rangle \sim t^{1/2}$.  
For a few cases, at low volume fractions, the crossover to the characteristic subdiffusive scaling regime occurs outside the time range shown in Fig.~\ref{figure_3}(a), which we have cropped for visibility. 
However, we confirmed via simulation all values of Pe and $\phi$ exhibit subdiffusive scaling at sufficiently long times. 
A general trend observed in Fig.~\ref{figure_3}(a) is that the overall magnitude of the MSD increases with $\text{Pe}$, and decreases with $\phi$.  
In addition, the crossover between the different scaling regimes exhibits a nontrivial dependence on $\text{Pe}$ and $\phi$, which we discuss in the discussion. 

{\color{black} A more detailed analysis of Fig.~\ref{figure_3}(a) reveals two different scenarios in the MSD's behavior.  
When $\text{Pe}$ and $\phi$ are both sufficiently large, the MSD transitions rapidly from a ballistic regime to a subdiffusive regime.  
However, for certain values of $\text{Pe}$ and $\phi$, an intermediate regime is observed where $\langle x^2(t) \rangle \sim t^\alpha$ with a scaling exponent $\alpha$ between $ \frac{1}{2} < \alpha < 2$. 
This intermediate scaling regime is typically observed when Pe or $\phi$ are small, and collisions with neighboring particles are infrequent.
Figure~\ref{figure_3}(b) provides a state diagram for the SF-ABP system and a simple scaling argument that approximately delineates these two MSD scenarios.}  
The solid line in Fig.~\ref{figure_3}(b) corresponds to the condition where the particle run length $\ell_0$ equals the mean free path $\lambda$.  
For values of $\text{Pe}$ and $\phi$ above this line, the MSD appears to transition directly from ballistic to subdiffusive behavior after a time $\tau_R$.  
Below this line, an intermediate regime, as described above, is observed.  
Here, we estimate the mean free path as the average distance between the edges of adjacent clusters.
In prior work~\cite{Schiltz-Rouse2023-vj}, we showed the mean free path is well-approximated by $\lambda_{ABP} = (1 - \phi)/(\rho \mathcal{T}_k^\beta) = \lambda/ \mathcal{T}_k^{\beta}$, where $\beta \approx 0.75$ and $\lambda$ is the mean free path of the passive SF system.  
Subdiffusive behavior in single-file systems is largely attributed to caging effects caused by collisions with neighboring particles~\cite{Ooshida2016-qu, Wittmann2021-nt}.  
When $\ell_0/\lambda_{ABP} > 1$, particles move ballistically between collisions with their neighbors, preventing any diffusive behavior in the MSD.  
In contrast, when $\ell_0/\lambda_{ABP} < 1$, particles exhibit more diffusive motion between collisions.  
A similar result has been observed in studies of passive single-file particles, where the onset of subdiffusive behavior typically arises when $t \approx \lambda^2/D$~\cite{Rallison_1988, Sane2010-sy, Schweers2023-eq}. 
In the discussion, we calculate the local exponent of the MSD, which provides a more quantitative picture of the crossover dependence of the various scaling regimes. 

In Fig.~\ref{figure_3}(c), we show the 1D-mobility $F$ measured from simulation, alongside the analytical result for the 1D-mobility of the passive hard particle system [Eq.~(\ref{eq:2})].
{\color{black} The 1D-mobility $F$ is measured by fitting a horizontal line to the long-time behavior of $\langle \Delta x^2 \rangle / (2 \sigma^2 t^{1/2})$.
Simulations are run for sufficient time, such that the long-time behavior of this quantity plateaus to a constant.}
As expected, in the limit $\text{Pe} \rightarrow 0$, the 1D-mobility converges to the analytical result for the passive system.  
For all values of $\text{Pe}$, the 1D-mobility decreases monotonically with increasing packing fraction.  
At higher packing fractions, particles experience more frequent collisions with neighboring particles, reducing mobility as crowding effects intensify.  
In contrast, for fixed $\phi$, the 1D-mobility increases with increasing $\text{Pe}$, reflecting the impact of self-propulsion on particle dynamics.  
As $\text{Pe}$ increases, particles move more persistently, allowing them to overcome some of the crowding effects imposed by neighboring particles, thus increasing their effective mobility. 

In Fig.~\ref{figure_3}(d), the normalized velocity autocorrelation function is shown for the representative selection of $\text{Pe}$ and $\phi$.  
Across all values of $\text{Pe}$ and $\phi$, the autocorrelation function exhibits self-similar behavior, characterized by an initial exponential decay, followed by a negative minimum between $\tau_R < t < 10 \, \tau_R$, and a subsequent power-law decay back to zero.  
The relaxation time of the normalized velocity autocorrelation function, relative to the ideal case, decreases with increasing $\text{Pe}$ and $\phi$, indicating faster decorrelation of particle velocities in systems with higher activity or packing fraction.  
The normalized velocity autocorrelation for the ideal case is plotted in Fig.~\ref{figure_3}(d) for reference (dashed black line).  
\begin{figure*}[t!]
	\centering
	\includegraphics[width=0.70\textwidth]{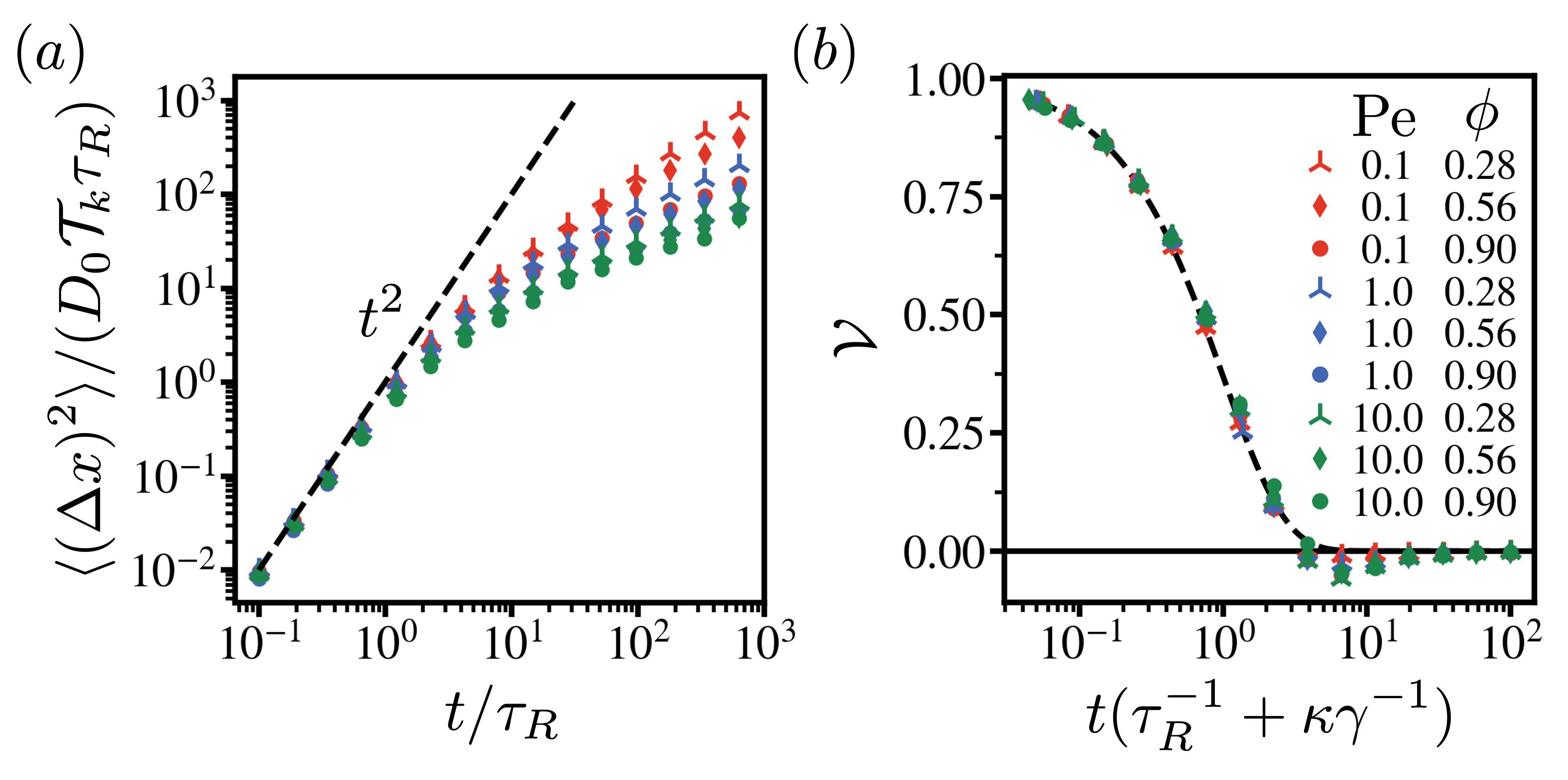}        
	\caption{\protect\small{(a) Universal collapse of the ballistic regime of the MSD when appropriately rescaled by the kinetic temperature [See Eq.~(\ref{eq:12})]. (b) Short-time collapse of the normalized velocity autocorrelation function, where the {\color{black} dashed black curve} corresponds to the analytical result $\mathcal{V}(t) = e^{-t(\tau_R^{-1} + \kappa \gamma^{-1})}$.}}
	\label{figure_4}
\end{figure*}
The power-law decay of the velocity autocorrelation function, which has characteristic scaling $-t^{-3/2}$ [see inset of Fig.~\ref{figure_3}(d)], has been widely observed in earlier simulations and experiments of passive systems~\cite{Tripathi2010-zt}.  
In passive single-file systems, this decay arises from long-lasting correlations imposed by caging effects.  
This power-law behavior, or long-time tail, has been extensively studied and is understood as a universal feature of confined, single-file passive systems~\cite{Jepsen1965-qg, Dekker1982-xj, Hagen1997-bo, Stepisnik2000-ku}.  
To date, this power-law decay has not been reported for active systems.  
However, this behavior is not entirely unexpected, as prior work on single-file active systems has observed the characteristic subdiffusive behavior (\(\langle (\Delta x)^2 \rangle \sim t^{1/2}\))~\cite{Bertrand2018-ll, Dolai2020-cu, Caprini2020-gj}, which suggests that the velocity autocorrelation function must decay as $-t^{-3/2}$.

\section{Discussion}

In principle, the key quantity required to calculate the MSD is the velocity autocorrelation function, or equivalently, the normalized velocity autocorrelation function $\mathcal{V}(\tau)$, and the mean square velocity $\langle v^2 \rangle$, as expressed in Eq.~(\ref{eq:9}).  
Even in the case of passive systems, it is a challenge to derive an analytical expression for the velocity autocorrelation function that is quantitative across all time scales for interacting many-body systems~\cite{mcquarrie2000statistical, berne2013dynamic}.  
We direct the reader to various empirical and first-principle approaches proposed for passive systems to capture features of the velocity autocorrelation function~\cite{boon1991molecular}.  
However, even without full knowledge of the normalized velocity autocorrelation function for the interacting SF-ABP system, a quantitative understanding of the MSD's short-time behavior can be obtained.  
As the mean square velocity is related to the reduced kinetic temperature via $\langle v^2 \rangle = \frac{1}{2} U_a^2 \mathcal{T}_k$, substituting this expression into Eq.~(\ref{eq:9}) gives:
\begin{equation}
    \label{eq:11}
    \langle (\Delta x)^2 \rangle = U_a^2 \mathcal{T}_k t \left[ \int_0^t \left( 1 - \frac{\tau}{t} \right) \mathcal{V}(\tau) d\tau \right] \, .
\end{equation}
A zeroth-order approximation for the normalized velocity autocorrelation function (i.e., $\mathcal{V}(\tau) = 1$) and subsequent integration of Eq.~(\ref{eq:11}), reveals that for all packing fractions and P\'eclet numbers, the motion of the particle is ballistic at short times, scaling as 
\begin{equation}
    \label{eq:12}
    \langle (\Delta x)^2 \rangle \sim \frac{1}{2} U_a^2 \mathcal{T}_k \, t^2 = \tau_R^{-1} D_0 \mathcal{T}_k \, t^2 \, .
\end{equation}
In Fig.~\ref{figure_4}(a), we show that the short-time ballistic behavior of the MSD collapses onto a universal curve by appropriately rescaling by $\mathcal{T}_k$.
Here, we arrive at a key observation: the MSD of a tagged ABP in the single-file system is always ballistic for times less than $\tau_R$, and the reduced kinetic temperature can quantitatively capture the particle's speed in this regime. 

In an attempt to capture the short-time behavior of the normalized velocity autocorrelation function, we approximate the collisional force $F_c$ to first order as $F_c = - \kappa \Delta x$, where $\kappa$ represents a generalized spring constant.  
We envision that the presence of neighboring particles induces a caging effect, which we approximate by a harmonic potential.  
Within this harmonic approximation for $F_c$, the normalized velocity autocorrelation function can be expressed as:  
\begin{equation}
    \label{eq:13}
    \mathcal{V}(t) = \frac{1}{1 - \tau_R \kappa \gamma^{-1}} \left(e^{-t/\tau_R} - e^{-t\kappa/\gamma}\right) + e^{-t\kappa/\gamma} \, .
\end{equation}
Focusing on the short-time behavior, we expand Eq.~(\ref{eq:13}) to first order, obtaining~$\mathcal{V}(t) = 1 - (\tau_R^{-1} + \kappa \gamma^{-1}) t + \mathcal{O}(t^2) \approx e^{-t(\tau_R^{-1} + \kappa \gamma^{-1})}$.  
For the functional form of the generalized spring constant, we construct a heuristic approximation $\kappa \sim \gamma U_a \sqrt{\mathcal{T}_k}/\lambda_{ABP}$, by identifying the relevant force scale acting on a tagged particle, $\gamma U_a \sqrt{\mathcal{T}_k}$, and the length scale over which the force acts given by the mean free path $\lambda_{ABP}$.  
The functional form of the generalized spring constant $\kappa$ captures the following trends:  
As the packing fraction increases, $\kappa$ increases, reflecting a more pronounced caging effect at higher packing fractions where particle collisions are more frequent.  
Conversely, as the active P\'eclet number increases at a fixed packing fraction, $\kappa$ decreases, indicating that particles with greater persistence overcome the caging effect from their neighbors more easily.  
In Fig.~\ref{figure_4}(b), we demonstrate that the short-time behavior of the normalized velocity autocorrelation function is well captured by $\mathcal{V}(t) = e^{-t(\tau_R^{-1} + \kappa \gamma^{-1})}$, successfully reflecting the key trends in activity and packing fraction at short times. 
This form for the normalized velocity autocorrelation function in Eq.~(\ref{eq:11}) provides a first-order correction to the MSD at short times. 
However, this expression does not capture the long-time behavior, particularly the power-law decay observed in the velocity autocorrelation function at later times.  

\begin{figure*}[t!]
	\includegraphics[width=0.95\textwidth]{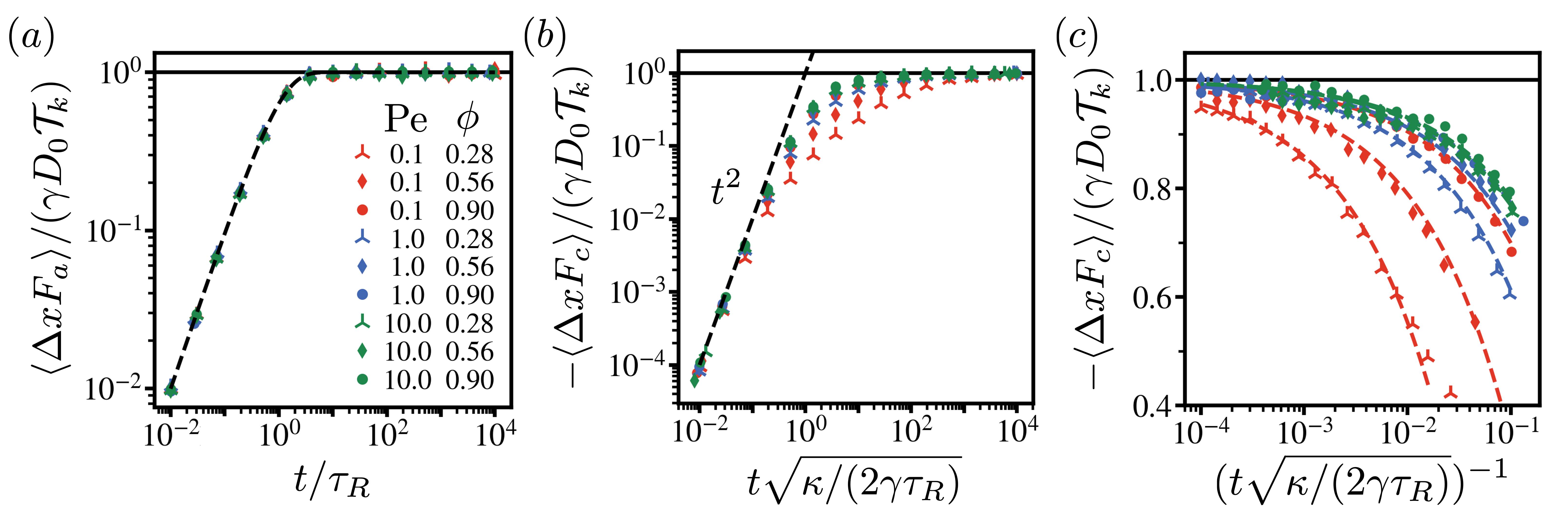}        
	\caption{\protect\small{(a) Correlation $\langle \Delta xF_a \rangle$ and (b) $\langle \Delta xF_c \rangle$ {\color{black} computed for the SF-ABP system} for the representative set of $\text{Pe}$ and $\phi$. The dashed black line in (a) corresponds to the analytical solution for $\langle \Delta xF_a \rangle$ [Eq.~(\ref{eq:21})]. In (b), the dashed black line corresponds to the short-time scaling. The long-time behavior (c) is illustrated using colored lines which correspond to Eq.~(\ref{eq:22}).} }
	\label{figure_5}
\end{figure*}

To further rationalize the observed trends in the MSD, we introduce the corresponding Smoluchowski equation, which governs the time evolution of the probability density function (PDF) for a tagged ABP's position and orientation. 
This conservation equation, which is formally equivalent to the equations of motion given in Eq.~(\ref{eq:4}), is expressed as:
\begin{equation}
    \label{eq:14}
    \partial_t P = -U_a \cos \theta \, \partial_x P - \frac{1}{\gamma} \partial_x (F_c P) + \frac{1}{\tau_R} \partial_{\theta \theta} P \, ,
\end{equation}
where $P = P(x, \theta, t | \, x_0, \theta_0, t_0)$ is the probability of finding a tagged ABP at position $x$ and orientation $\theta$ at time $t$, given that it was at position $x_0$ and orientation $\theta_0$ at an earlier time $t_0$.
The PDF satisfies the normalization condition $\int dx \, d\theta \, P = 1$, and we assume an initial condition of $P(x_0, \theta_0, t_0) = \delta(x - x_0) / (2\pi)$, where the ABP is initially localized at $x_0$ with a uniform distribution in its orientation. 
Without loss of generality, we set $x_0 = 0$ and $t_0 = 0$.
The ensemble average of any microscopic observable $f[x(t), \theta(t)]$ is then defined as:
\[
\langle f[x(t), \theta(t)] \rangle = \int dx \, d\theta \, f(x, \theta) P(x, \theta, t | \, x_0, \theta_0, t_0) \, .
\]

Following a similar approach to Schakenraad et al.~\cite{Schakenraad2020-mj}, we express the time derivative of the MSD as:
\begin{equation}
    \label{eq:15}
    \partial_t\langle (\Delta x)^2 \rangle = \int dx \, d\theta \, (\Delta x)^2 \, \partial_t P \, ,
\end{equation}
which, upon substituting Eq.~(\ref{eq:14}) into the right-hand side of Eq.~(\ref{eq:15}), gives:
\begin{equation}
    \label{eq:16}
    \partial_t \langle (\Delta x)^2 \rangle = \frac{2}{\gamma} \bigg[ \langle \Delta x F_a \rangle + \langle \Delta x F_c \rangle \bigg] \, ,
\end{equation}
where the time derivative of the MSD is expressed in terms of two correlations: $\langle \Delta x F_a \rangle$ and $\langle \Delta x F_c \rangle$.  
Interestingly, we show that the short-time ballistic behavior of the MSD is primarily governed by the correlation $\langle \Delta x F_a \rangle$ and the long-time subdiffusive behavior is due to $\langle \Delta x F_c \rangle$.
Additionally, in the asymptotic limit, the left-hand side of Eq.~(\ref{eq:16}) vanishes as 
\begin{equation*}
\lim\limits_{t \rightarrow \infty} \partial_t \langle (\Delta x)^2 \rangle \sim t^{-\frac{1}{2}} \rightarrow 0
\end{equation*}
establishing the long-time relationship: 
\begin{equation}
    \label{eq:17}
    \lim\limits_{t \rightarrow \infty} \langle \Delta x F_a \rangle = \lim\limits_{t \rightarrow \infty} - \langle \Delta x F_c \rangle \, .
\end{equation}

We proceed by deriving analytical expressions for $\langle \Delta x F_a \rangle$ and $\langle \Delta x F_c \rangle$, which allow us to obtain the MSD by integrating Eq.~(\ref{eq:16}).
We first express the time derivative of $\langle \Delta xF_a \rangle$ as
\begin{equation}
    \label{eq:18}
    \partial_t \langle \Delta xF_a \rangle = \frac{\gamma U_a^2}{2} + \frac{1}{\gamma}\langle F_aF_c \rangle - \frac{1}{\tau_R} \langle \Delta xF_a \rangle \,
\end{equation}
using the same approach to derive Eq.~(\ref{eq:16}).
The correlation $\langle F_aF_c \rangle$, which is time independent, can be written in terms of the reduced kinetic temperature as 
\begin{equation}
    \label{eq:19}
    \langle F_aF_c \rangle = -\langle F_c^2 \rangle = -\frac{(\gamma U_a)^2}{2} \left(1 - \mathcal{T}_k \right) \, ,
\end{equation}
where we have used Eq.~(\ref{eq:5}) and the unique property of ABPs $\langle vF_c \rangle = 0$~\cite{Schiltz-Rouse2023-vj}.
Upon substituting Eq.~(\ref{eq:19}) into Eq.~(\ref{eq:18}), we obtain the first-order differential equation
\begin{equation}
    \label{eq:20}
    \partial_t \langle \Delta xF_a \rangle = \frac{\gamma U_a^2}{2} \mathcal{T}_k - \frac{1}{\tau_R} \langle \Delta xF_a \rangle \, ,
\end{equation}
with solution 
\begin{equation}
    \label{eq:21}
    \langle \Delta xF_a \rangle =  \gamma D_0 \mathcal{T}_k \left[ 1 - e^{-t/\tau_R} \right] \, .
\end{equation}
In Fig.~\ref{figure_5}(a), we show excellent agreement between Eq.~(\ref{eq:21}) and simulation results demonstrating that all $\phi$ and $\text{Pe}$ values collapse onto a universal curve.
By integrating Eq.~(\ref{eq:21}), we obtain the predominant short-time contribution to the MSD. 
We emphasize that Eq.~(\ref{eq:21}) can be trivially extended to ABPs in higher dimensions. 
However, in these higher dimensional systems, the $\text{Pe}$ and $\phi$ dependence of the reduced kinetic temperature has not been fully quantified.  

The correlation $\langle \Delta xF_c \rangle$ presents a greater challenge to calculate analytically due to the spatiotemporal complexity of the interparticle force $F_c$.  
Rather than providing an exact treatment, we perform a scaling analysis for both the short- and long-time behavior, which is sufficient to obtain an accurate expression for the MSD.  
Figure~\ref{figure_5}(b) presents simulation results for $\langle \Delta xF_c \rangle$ over the representative range of $\phi$ and $\text{Pe}$.  
At short times, all curves exhibit similar scaling behavior that is captured by our harmonic approximation for $F_c$, which gives the correct scaling $\langle \Delta x F_c \rangle = -\kappa \langle (\Delta x)^2 \rangle \sim - \kappa D_0\mathcal{T}_k t^2/\tau_R$ for times less than $\tau_R$.  
The short-time behavior collapses onto a universal curve when scaled by the timescale $\sqrt{(2 \gamma \tau_R)/\kappa}$, as shown in Fig.~\ref{figure_5}(b) (see Supplementary Material for calculation details).  
The quadratic scaling of $\langle \Delta xF_c \rangle$ at short times confirms that its contribution to the MSD is nominal.
At short times, $\langle \Delta xF_a \rangle$ contributes a ballistic term to the MSD, which grows more rapidly in time than the contribution from $\langle \Delta xF_c \rangle$, which scales as $\langle (\Delta x)^2 \rangle \sim t^3$.
At long times, $\langle \Delta xF_c \rangle$ plateaus to $-\gamma D_0 \mathcal{T}_k$, consistent with the asymptotic relationships derived in Eqs.~(\ref{eq:17}) and (\ref{eq:21}), where 
\begin{equation*}
\lim\limits_{t \rightarrow \infty}\langle \Delta xF_c \rangle =  - \lim\limits_{t \rightarrow \infty} \langle \Delta xF_a \rangle = - \gamma D_0 \mathcal{T}_k \, .
\end{equation*}

\begin{figure*}[t!]
    \includegraphics[width=0.70\textwidth]{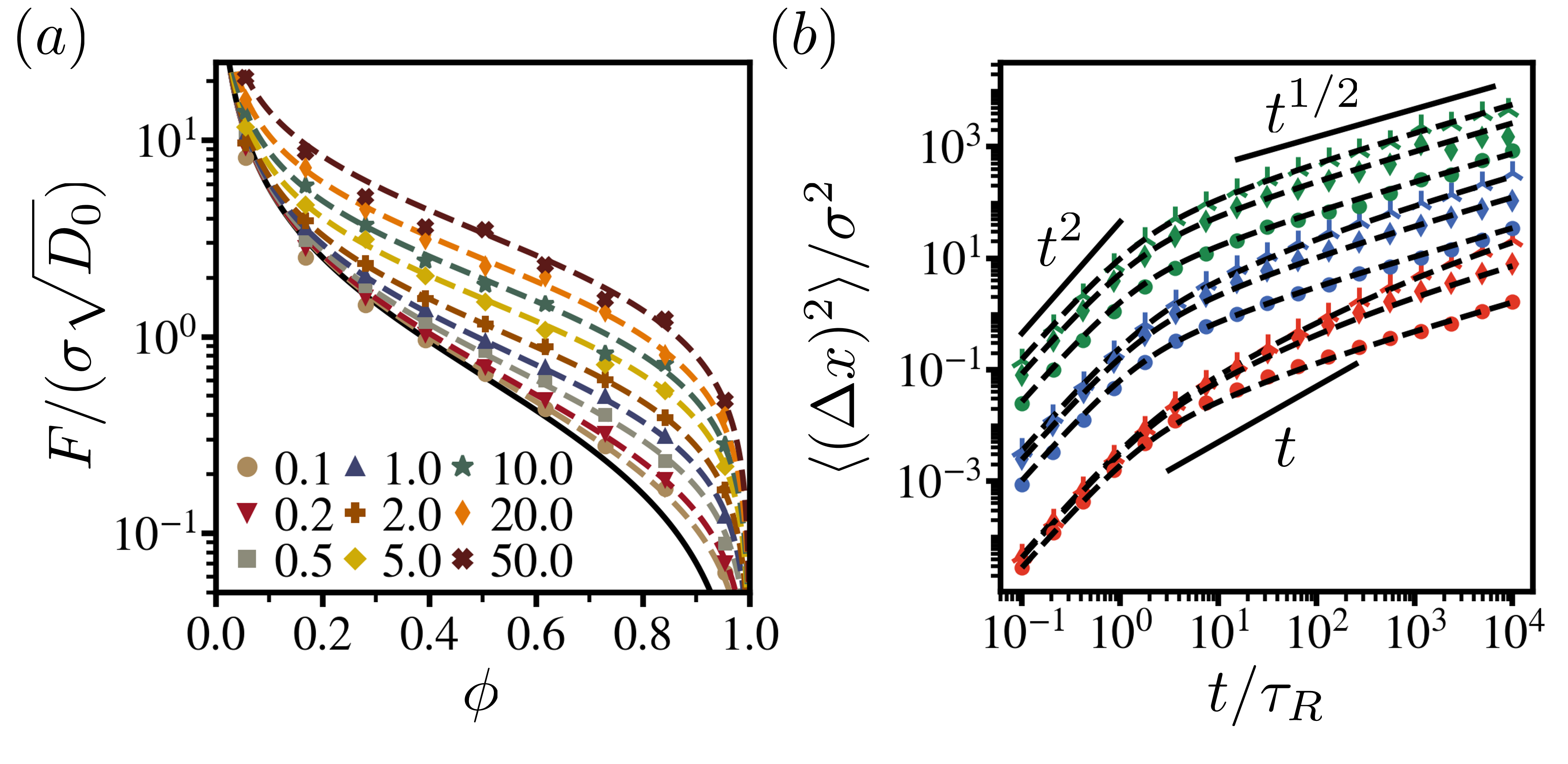}        
    \caption{\protect\small{(a) 1D-mobility for the SF-ABP system as a function of packing fraction for different active P\'eclet numbers. Simulation results are plotted as points and the analytical expression, Eq.~(\ref{eq:23}), is plotted as a dashed colored line. {\color{black}Symbols and colors refer to different values of Pe}. (b) Scaled MSD measurements of a tracer particle in the SF-ABP system for different active P\'eclet numbers and packing fractions. Simulation results are plotted using colored points and the analytical expression, Eq.~(\ref{eq:27}), is shown using black dashed lines.}}
    \label{figure_6}
\end{figure*}

To bridge the short- and long-time regimes, we consider the limit $t\gg \tau_R$ and estimate the intermediate-time behavior of $\langle \Delta xF_c \rangle$ from Eq.~(\ref{eq:16}) as
\begin{align}
    \label{eq:22}
    \langle \Delta xF_c \rangle &  =
    \left[\frac{\gamma}{2} \partial_t \langle (\Delta x)^2 \rangle   - \langle \Delta x F_a \rangle \right]
    \nonumber
    \\
    & \sim  \frac{\gamma F}{2 t^{1/2}} - \gamma D_0 \mathcal{T}_k \, .    
\end{align}
The second line of Eq.~(\ref{eq:22}) follows from two key observations.  
At sufficiently long times all packing fractions and P\'eclet numbers exhibit subdiffusive scaling, leading to $\partial_t \langle (\Delta x)^2 \rangle = F / t^{1/2}$. 
In the Supplementary Material, we include an analytical calculation demonstrating this fact using an active variant of the Rouse model. 
The second observation follows from Eq.~(\ref{eq:21}), where $\langle \Delta x F_a \rangle = \gamma D_0 \mathcal{T}_k$ for times larger than $\tau_R$.

Interestingly, the 1D-mobility of the SF-ABP system can be predicted using Kollmann's result for equilibrium single-file systems,
\begin{equation}
    \label{eq:23}
    F = \frac{1}{\rho}\sqrt{\frac{D_0 \mathcal{X}}{\pi}} \, ,
\end{equation}
where the only modification is that $\mathcal{X}$ is given by the constant Pe compressibility [Eq.~(\ref{eq:8})], and $D_0 = \frac{1}{2} U_a^2 \tau_R$, the free diffusion coefficient of an ideal ABP.  
In Fig.~\ref{figure_6}(a), we demonstrate excellent agreement between the measured 1D-mobility from simulation and Eq.~(\ref{eq:23}).  
Using this expression for the 1D-mobility, we show in Fig.~\ref{figure_5}(c) good agreement between the intermediate-time scaling of $\langle \Delta xF_c \rangle$ [Eq.~(\ref{eq:22})] and simulation results.

Although derived for equilibrium Brownian particles, Kollmann's result for the 1D-mobility appears to apply to a broader class of particle dynamics.  
Kollmann's core insight was to relate a tagged particle's MSD in the long-time limit to the static structure factor, which can, in turn, be related to the system's isothermal compressibility~\cite{Pusey1985-xa}.  
Since the static structure factor is a purely mechanical quantity depending only on particle positions, this connection remains valid for non-equilibrium single-file systems, as long as the static structure factor remains well-defined.  
For ABPs, prior work has shown that the relationship between the static structure factor and the constant Pe compressibility mirrors that in equilibrium systems~\cite{Dulaney2021-kf}.
These observations validate the extension of Kollmann's result to the 1D-mobility of ABPs. 
Further work is needed to address the scope of this result and determine whether it can be extended to an even more general class of non-equilibrium particle dynamics.

To construct an accurate expression for the MSD of the SF-ABP system, we adopt an approach similar to Lin et al.~\cite{Lin2005-uh}, where the ansatz for the MSD is expressed as
\begin{equation}
    \label{eq:24}
    \frac{1}{\langle (\Delta x)^2 \rangle} = \frac{1}{\langle (\Delta x)^{2}\rangle_{\text{short}}} + \frac{1}{\langle (\Delta x)^{2}\rangle_{\text{long}}} \, .
\end{equation}
This form of the MSD [Eq.~(\ref{eq:24})] captures the short- and long-time behavior of the MSD and smoothly interpolates between these two regimes at intermediate times.  
We model the short-time contribution to the MSD as
\begin{multline}
    \label{eq:25}
    \langle (\Delta x)^2 \rangle_{\text{short}} = \frac{2}{\gamma}\int_0^t dt' \langle \Delta x F_a \rangle \\
    = 2 D_0 \mathcal{T}_k \bigg[t+\tau_{R}\left(e^{-t/\tau_{R}}-1\right)\bigg] \, ,
\end{multline}
where $\langle \Delta x F_a \rangle$ is given by Eq.~(\ref{eq:21}).  
This contribution is identical to the MSD of an ideal ABP [Eq.~(\ref{eq:10})], except for the prefactor $\mathcal{T}_k$, which accounts for the reduction in particle speeds due to collisions.
At short times, Eq.~(\ref{eq:24}) predicts ballistic motion, scaling as $\langle (\Delta x)^2 \rangle \sim \frac{1}{2}U_a^2 \mathcal{T}_k t^2$, consistent with both simulation results and the short-time expansion in Eq.~(\ref{eq:12}).  
The long-time contribution to the MSD is given by
\begin{align}
    \label{eq:26}
    \langle (\Delta x)^2 \rangle_{\text{long}} &= \frac{2}{\gamma}\int_0^t dt' \left( \langle \Delta x F_c \rangle + \langle \Delta x F_a \rangle \right) \notag \\
    &= \frac{2}{\gamma}\int_0^t dt' \left( \frac{\gamma F}{2 \sqrt{t'}} - \gamma D_0 \mathcal{T}_k + \gamma D_0 \mathcal{T}_k \right) \notag \\
    &= 2 F t^{1/2}\, ,
\end{align}  
where $\langle \Delta x F_c \rangle$ is given by Eq.~(\ref{eq:22}) and $\langle \Delta x F_a \rangle$ reaches its asymptotic value of $\gamma D_0 \mathcal{T}_k$.  
Essentially, at long times, it is only necessary to retain the subdiffusive contribution to the MSD.  
Combining the short- and long-time behaviors, we arrive at the full expression of the MSD:
\begin{widetext}
\begin{equation}
    \label{eq:27}
    \langle (\Delta x)^{2} \rangle 
    = \dfrac{2 D_0 \mathcal{T}_k \left[t + \tau_R \left(e^{-t/\tau_R} - 1\right)\right]}{1 + D_0 \mathcal{T}_k \left[t + \tau_R \left(e^{-t/\tau_R} - 1\right)\right] \dfrac{\phi}{\sigma_p} \sqrt{\dfrac{\pi}{D_0 \mathcal{X} t}}} \, .
\end{equation}
\end{widetext}
In Fig.~\ref{figure_6}(b), we compare Eq.~(\ref{eq:27}) and simulation results for the selected values of $\text{Pe}$ and $\phi$ and find excellent agreement across all time scales.  
By design, the ansatz for the MSD captures the short- and long-time behavior, so it is not surprising we show good agreement in these regimes, as our scaling analysis accurately predicts these regimes. 
We find the performance of Eq.~(\ref{eq:27}) at intermediate times more surprising, where
the intrinsic interpolation from the MSD ansatz accurately captures the crossover between the ballistic and subdiffusive regime. 

\begin{figure*}[t!]
    \includegraphics[width=0.70\textwidth]{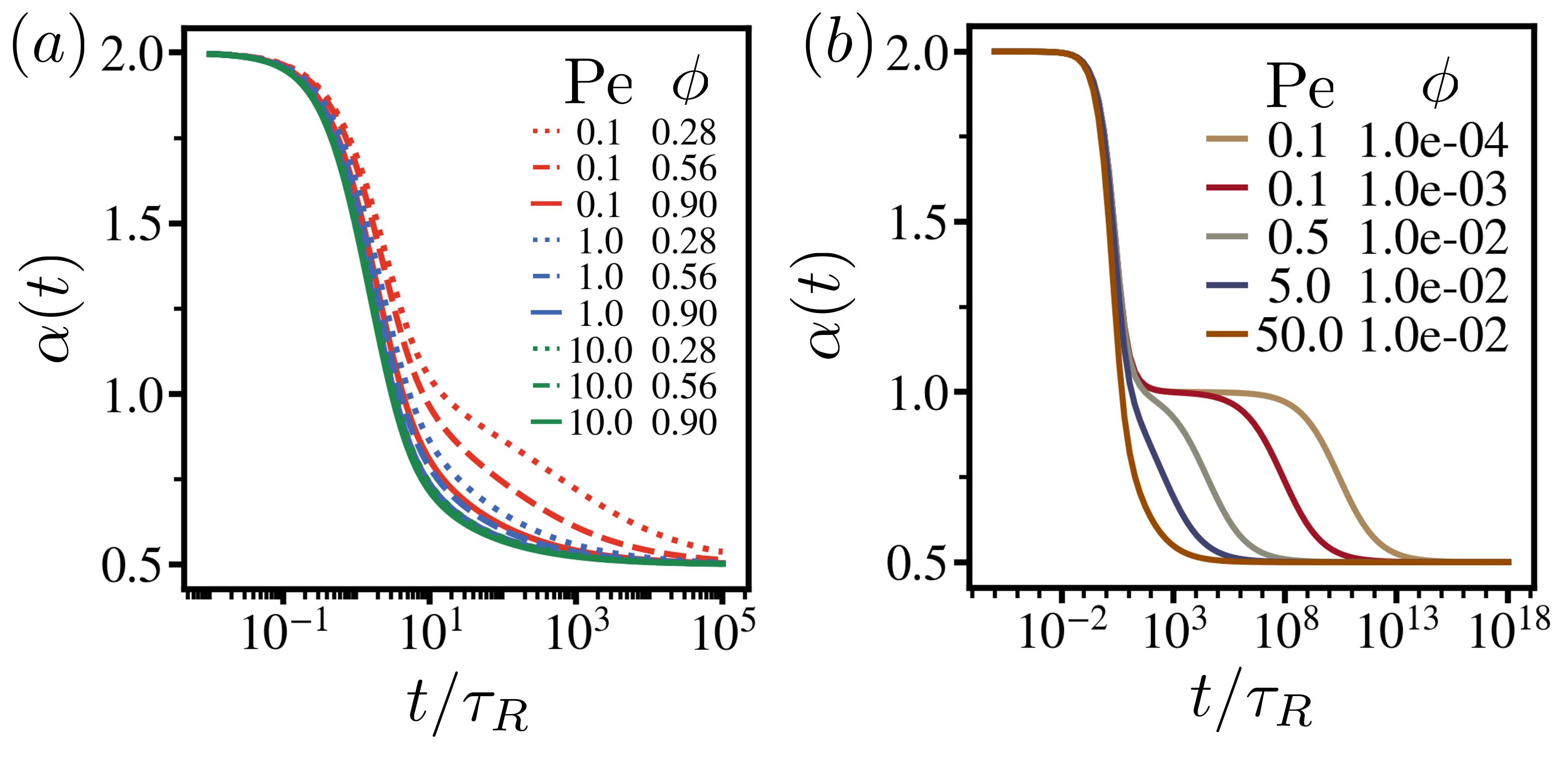}        
    \caption{\protect\small{{\color{black}(a) Local exponent from Eq.~(\ref{eq:27}) as a function of rescaled time for the representative set of Pe and $\phi$. The local exponent initially has a value of 2 representing the ballistic regime in the MSD. The emergence of a diffusive regime at intermediate times is illustrated in (b) for large values of Pe. At long times, subdiffusive scaling is recovered for all values of Pe and $\phi$, illustrated by the local exponent's value of 0.5. (b) The local exponent for additional values of $\text{Pe}$ and $\phi$. We have selected values that better highlight the diffusive regime at intermediate times. Note that the duration of this regime is highly sensitive to the specific values of $\text{Pe}$ and $\phi$. However, it is possible to identify values of $\text{Pe}$ and $\phi$ that will give a desired profile for the different scaling regimes as well as tune the overall magnitude of the MSD.}}}
    \label{figure_7}
\end{figure*}

In Fig.~\ref{figure_7}(a), we calculate the local exponent $\alpha = d \ln (\langle \Delta x^2 \rangle)/d \ln(t)$, which quantifies the scaling exponent of the MSD as a function of time (i.e., $\langle \Delta x^2 \rangle \sim t ^\alpha$).
At short times $\alpha \approx 2$ for all $\text{Pe}$ and $\phi$, illustrating that the MSD is ballistic at short times.
After times greater than approximately $\tau_R$, the local exponent decreases with a scaling that depends on $\text{Pe}$ and $\phi$.
The trends observed for the local exponent agree with the simple scaling arguments discussed in Fig.~\ref{figure_3}(b).
For sufficiently large $\phi$ and $\text{Pe}$ values, the local exponent rapidly decays from $\alpha = 2$ to $\alpha = \frac{1}{2}$, characteristic of the MSD transitioning directly from ballistic to subdiffusive behavior. 
For the case where the MSD exhibits diffusive character at intermediate times (i.e. when $\phi$ and $\text{Pe}$ are sufficiently small), the local exponent rapidly decreases from $\alpha = 2$ to $\alpha =1$ within a time $\tau_R$, followed by a slower decay that can expand several decades to reach the asymptotic value of $\alpha = \frac{1}{2}$.
A short-time expansion of the local exponent for the MSD shows the following short-time scaling behavior
\begin{equation}
\alpha (t) = 2 - \frac{t}{3\tau_R} - \frac{3\text{Pe}\mathcal{T}_{k}\phi\sqrt{\pi t^3} }{4\sqrt{2\mathcal{X} \tau_R^3} } + \mathcal{O}(t^{2}) \, .
\end{equation}
The second term of this expansion corresponds to the decay of $\alpha$ driven by the intrinsic reorientation of the particle via rotational diffusion. In contrast, the third term can be associated with further reduction in $\alpha$ due to collisions with neighboring particles which depends on both $\text{Pe}$ and $\phi$. 
In Fig.~\ref{figure_7}(b), we include an additional plot for the local exponent of the MSD for different values of $\text{Pe}$ and $\phi$, illustrating how the crossover between various scaling regimes can be controlled. 
From an engineering and applications perspective, Eq.~(\ref{eq:27}) encapsulates a design rule for obtaining a desired profile for the MSD of a tagged ABP, where it is possible to tune both the overall magnitude of the MSD and the duration of the various scaling regimes. 
 
\section{Conclusion}  

This study characterizes the single-file diffusion (SFD) of active Brownian particles (ABPs) using a combination of Brownian dynamics simulations and analytical theory.  
Using scaling and heuristic arguments, we derived an accurate analytical expression for the mean square displacement (MSD) of a tagged ABP [Eq.~(\ref{eq:27})] as a function of $\text{Pe}$ and $\phi$.  
Additionally, we demonstrated that Kollmann’s 1D-mobility expression [Eq.~(\ref{eq:23})], initially developed for equilibrium systems, can be extended to active systems with minimal modification, highlighting key similarities between passive and active single-file systems.  
While our analytical model effectively captures the essential dynamics, open questions remain.  
Future work could involve solving the correlation term $\langle \Delta x F_c \rangle$ exactly, which may offer modest improvements in predicting the MSD.  
Moreover, finding an exact analytical solution to the governing Smoluchowski equation [Eq.~(\ref{eq:14})], if feasible, would allow for precise calculations of higher-order moments, further refining our understanding of positional and orientational dynamics in SFD systems.

Future studies will explore systems with more complex interparticle interactions {\color{black} and the effects of external forces~\cite{Straube2024-fu}}, potentially revealing new scaling behaviors or emergent transport mechanisms.
An important open question is how these results differ when the particles are capable of inducing interparticle torques on one another such as in magnetic or patchy particle systems.
In addition, hydrodynamic interactions and specific self-propulsion mechanisms may significantly influence these findings.
{\color{black} Lastly, in prior work on passive systems, it has been shown that inertial effects can modify the scaling of the MSD and, in some cases, eliminate subdiffusive behavior at long times~\cite{Kollmann2003-as, Mon2003-ei}. 
It would be interesting to investigate the role of inertia in the context of active colloids under single-file confinement}

Another natural extension of this study involves investigating how the MSD changes as a function of the channel width. 
In passive systems, there has been shown to be a crossover from single-file to Fickian diffusion at a characteristic channel width~\cite{Sane2010-sy}.  
Such a study would systematically elucidate the differences between transport in passive and active systems under varying degrees of confinement.  
Such extensions have practical relevance for applications in confined environments, with implications for the design of advanced microfluidic systems, drug delivery platforms, and environmental sensing mechanisms.  
Finally, experimental studies of single-file active colloids could offer exciting opportunities to test these predictions and evaluate their practical impact.

\section{Supplementary Material}  

{\color{black}The supplementary material provides a detailed derivation of the short-time quadratic behavior of $\langle \Delta x F_c \rangle$ and the long-time subdiffusive behavior of the mean square displacement within the framework of the Active Rouse Model.}

%

\end{document}


\preprint{APS/123-QED}

\title{Supplemental Material: Single-File Diffusion of Active Brownian Particles}

\author{Akinlade Akintunde}
\affiliation{Department of Chemistry, The Pennsylvania State University, University Park, Pennsylvania, 16802, USA}

\author{Parvin Bayati}
\affiliation{Department of Chemistry, The Pennsylvania State University, University Park, Pennsylvania, 16802, USA}

\author{Hyeongjoo Row}
\affiliation{Department of Chemical and Biomolecular Engineering, UC Berkeley, Berkeley, CA 94720, USA}

\author{Stewart A. Mallory}
\email{sam7808@psu.edu}
\affiliation{Department of Chemistry, The Pennsylvania State University, University Park, Pennsylvania, 16802, USA}
\affiliation{Department of Chemical Engineering, The Pennsylvania State University, University Park, Pennsylvania, 16802, USA}


\date{\today}

\maketitle

\section{Short time quadratic behavior of $\langle \Delta xF_c \rangle$}

Within the harmonic approximation for the collisional force $(F_c = -\kappa \Delta x)$ and using the same technique to derive Eqs. (16) and (18) of the main text, we can express the time derivative of $\partial_t\langle \Delta xF_c\rangle$ as
\begin{equation}
    \partial_t \langle \Delta xF_c \rangle = - \kappa \, \partial_t \langle (\Delta x)^2 \rangle = - \frac{2 \kappa}{\gamma} \bigg[ \langle \Delta x F_a \rangle + \langle \Delta x F_c \rangle \bigg] \, ,
    \label{eq:S1}
\end{equation}
revealing an important timescale for the SF-ABP system, $\tau_H \equiv \gamma/\kappa$.
Using $e^{2t/\tau_{H}}$ as in integrating factor, we obtain
\begin{equation}
    \langle \Delta xF_c\rangle = -\frac{2}{\tau_H}\int_0^t \, d\tau \, \langle \Delta xF_a\rangle e^{-2(t-\tau)/\tau_H} \, ,
    \label{eq:S2}
\end{equation}
and substitution of $\langle \Delta xF_a\rangle$ from Eq. (21) of the main text gives
\begin{equation}
    \langle \Delta xF_c \rangle = \\ - \dfrac{\gamma D_0 \mathcal{T}_k}{1 - 2 \tau_R/\tau_H} \bigg[ 1 - e^{-2t/\tau_H} - 2\frac{\tau_R}{\tau_H} \left( 1 - e^{-t/\tau_R}\right)  \bigg] \, .
    \label{eq:S3}
\end{equation}
A short-time expansion for $t \ll \text{min}(\tau_H, \tau_R) $ results in $\langle \Delta xF_c \rangle \sim - \kappa D_0 \mathcal{T}_k \, t^2 / \tau_R$, illustrating the quadratic scaling at short times and the timescale $\sqrt{(2 \gamma \tau_R)/\kappa}$ used to obtain the collapse in Fig.~5(b) of the main text.

\section{Long time behavior of the mean square displacement (Active Rouse Model)}

\begin{figure}[h!]
    \centering
    \includegraphics[width=0.5\linewidth]{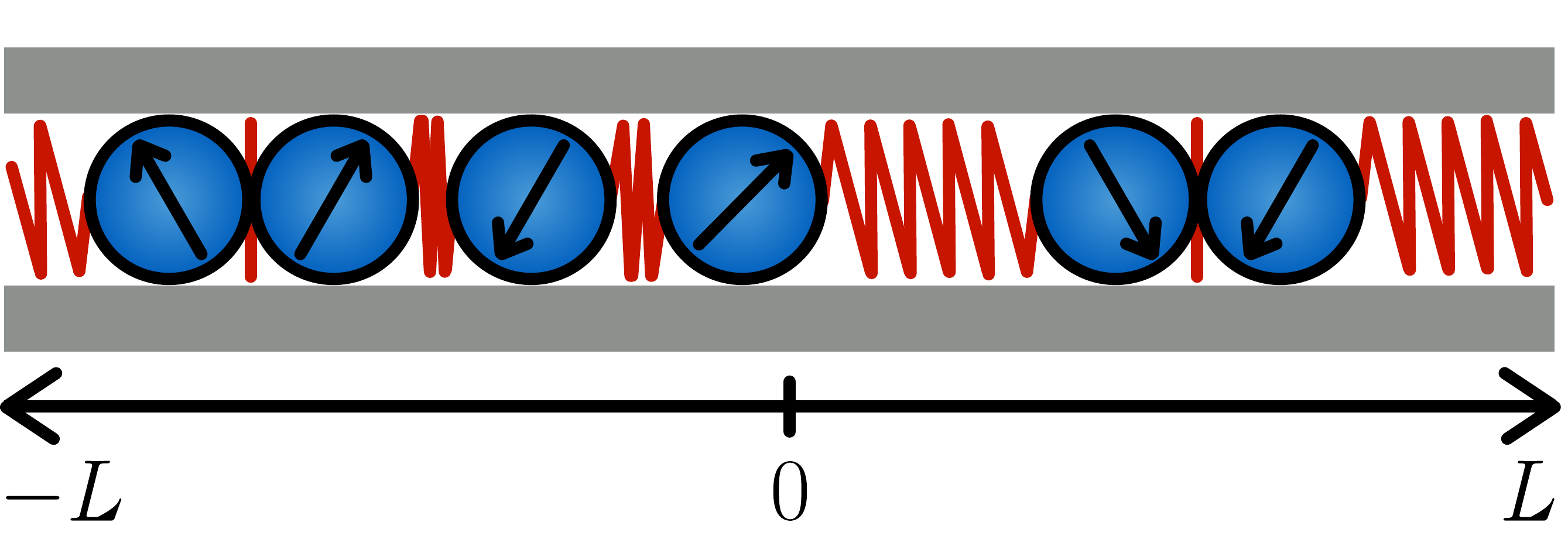}
    \caption{Schematic of active Rouse SF-ABP system where interparticle interactions are modeled as springs. Each purely repulsive active particle moves at a constant speed of $U_a \cos \theta$ while undergoing rotational Brownian motion with a reorientation time of $\tau_R$.}
    \label{fig:1}
\end{figure}

In this calculation, we model the SF-ABP system as an active variant of the Rouse Model, where active Brownian particles are bonded via harmonic springs in a one-dimensional geometry. A schematic illustrating the model is shown in Fig.~\ref{fig:1}.
Within the context of this model, the equation of motion of a tagged ABP or ``monomer" can be written as~\cite{Centres2010-qb, Leibovich2013-my, Lomholt2014-we},
\begin{equation}
    v(n,t) = \partial_{t}x(n,t) = \frac{1}{\gamma}F_{a}(n,t) + \frac{1}{\tau_H} [x(n+1,t) + x(n-1, t) - 2x(n,t)] \, ,
    \label{eq:S4}
\end{equation}
where $x(n,t)$ is the position of the tagged particle ($n^{th}$ particle) at time $t$ with its initial position set to zero, $x(n,t=0)=0$ and $\kappa$ is the spring constant of the harmonic interaction potential between adjacent ABPs. 
To match the geometry of the SF-ABP system studied in the main text, we consider a one-dimensional periodic system of length $2L$ where the first and last particles in the channel are bonded. 
We work in the limit where both the length of the channel and the number of particles are large, but the ratio $N/L$ remains finite.
In this limit, 
we apply a continuum approximation and perform the Fourier-Laplace transform of the particle position as,
\begin{equation}
\hat{\tilde{x}}(q,s) = \int_{-\infty}^{\infty} dn \int_{0}^{\infty} dt\, e^{-iqn}e^{-st} x(n,t) \, ,
    \label{eq:S5}
\end{equation}
with inverse transform,
\begin{equation}
   x(n,t)  = \frac{1}{4\pi^{2}i} \int_{-\infty}^{\infty} dq \int_{c-i\infty}^{c+i\infty} ds\, e^{iqn}e^{st} \hat{\tilde{x}}(q,s) \, ,
   \label{eq:S6}
\end{equation}
where $c$ is a real number greater than the real part of all singularities of $\hat{\tilde{x}}(q,s)$.
Using a second-order centered difference scheme and applying the Fourier-Laplace transform, Eq.~(\ref{eq:S1}) can be rewritten as 
\begin{equation}
    s\hat{\tilde{x}}(q,s) = \frac{1}{\gamma}\hat{\tilde{F}}_{a}(q,s) - \frac{q^2}{\tau_H}\hat{\tilde{x}}(q,s) \, .
    \label{eq:S7}
\end{equation}
From Eq.~(\ref{eq:S7}), we can compute the translational autocorrelation function of a particle in Fourier-Laplace space,
\begin{equation}
    \langle \hat{\tilde{x}}(q_{1},s_{1})\hat{\tilde{x}}(q_{2},s_{2})\rangle = \frac{1}{\gamma^2}\frac{\langle \hat{\tilde{F}}_{a}(q_{1},s_{1})\hat{\tilde{F}}_{a}(q_{2},s_{2})\rangle}{(s_{1}+q_{1}^{2}/\tau_H)(s_{2}+q_{2}^{2}/\tau_H)} \, ,
    \label{eq:S8}
\end{equation}
where $\langle \hat{\tilde{F}}_{a}(q_{1},s_{1})\hat{\tilde{F}}_{a}(q_{2},s_{2})\rangle$ is the Fourier-Laplace transform of the active force autocorrelation function~\cite{ten-Hagen2009-gt, graham2018microhydrodynamics},
\begin{equation}
\label{eq:S9}
\langle F_a (n_1,t_1)F_a(n_2,t_2)\rangle = \frac{1}{2}(\gamma U_a)^2e^{-|t_1-t_2|/\tau_{R}}\delta (n_{1}-n_{2}) \, .
\end{equation}
Here, the Dirac delta function $\delta(n_1-n_2)$ is a continuum analog of the Kronecker delta $\delta_{n_1 n_2}$, which indicates that active forces acting on different particles are uncorrelated.
In the time scales much longer than the reorientation time $\tau_R$, the autocorrelation of the active force
can be approximated as
\begin{equation}
\label{eq:S10}
\langle F_a (n_1,t_1)F_a(n_2,t_2)\rangle \approx \frac{\tau_R}{4}(\gamma U_a)^2 \delta(t_1 - t_2)\delta (n_{1}-n_{2}) \, .
\end{equation}
The Fourier-Laplace transform of $\langle F_a (n_1,t_1)F_a(n_2,t_2)\rangle$ gives
\begin{equation}
    \langle \hat{\tilde{F}}_{a}(q_{1},s_{1})\hat{\tilde{F}}_{a}(q_{2},s_{2})\rangle = \frac{\pi\tau_R (\gamma U_a)^2\delta (q_{2} + q_{1}) }{2(s_1+s_2)} \, .
    \label{eq:S11}
\end{equation}
Equation~(\ref{eq:S11}) can be substituted into Eq.~(\ref{eq:S8}) to obtain an expression for the positional autocorrelation function
\begin{equation}
    \langle \hat{\tilde{x}}(q_{1},s_{1})\hat{\tilde{x}}(q_{2},s_{2})\rangle = \frac{\tau_R\pi U_a^2\delta (q_{2} + q_{1}) }{2(s_1+s_2)(s_{1}+q_{1}^{2}/\tau_H)(s_{2}+q_{2}^{2}/\tau_H)} \, .
    \label{eq:S12}
\end{equation}
To invert the autocorrelation function to real space and time, we first perform the inverse Laplace transform
\begin{equation}
    \langle \tilde{x}(q_{1}, t_{1})\tilde{x}(q_{2}, t_{2})\rangle = -\frac{1}{4\pi^2}\int_{c-i\infty}^{c+i\infty}ds_{1}e^{s_1 t_1}\int_{c'-i\infty}^{c'+i\infty}ds_{2} e^{s_{2}t_{2}}\langle \hat{\tilde{x}}(q_{1},s_{1})\hat{\tilde{x}}(q_{2},s_{2})\rangle \, .
    \label{eq:S13}
\end{equation}
Both integrations here can be evaluated by rewriting the expression as a limit of contour integrals. 
We first calculate the integral over $s_2$ and assume without loss of generality that $s_2<s_1$ and $t_2<t_1$.
A Bromwich contour shown in Fig.~\ref{Fig:1} (assuming $t_2>0$) is used to evaluate the following contour integrals
\begin{equation}
\oint_{\Omega} dz F(z) = \int_{\Omega_1} dz F(z) + \int_{\Omega_2} dz F(z) = 2\pi i \sum_{b=1}^{2} \mathop{\mathrm{Res}}_{z = z_b}F(z) \, ,
\label{eq:S14}
\end{equation}
where the integrand $F(z) = e^{tz} \langle \hat{\tilde{x}}(q_{1},s_{1})\hat{\tilde{x}}(q_{2},z)\rangle $ has two simple poles at $z_{1}=-s_1$ and $z_{2}=-q^{2}_{2}/\tau_H$, and $\mathop{\mathrm{Res}}_{z = z_b} F(z)$ represents the residue of $F(z)$ at the pole $z=z_b$. 
The chosen contour $\Omega$ allows the integral over the curved arc $\Omega_{2}$ to vanish due to Jordans's lemma for $t_2 >0$.
Thus, the integral over $\Omega_1$ is of primary interest and equals the sum of the residues as follows
\begin{equation}
\int_{\Omega_{1}} dz F(z)
=\int_{c-i\infty}^{c+i\infty} ds_{2} e^{s_{2}t_{2}}\langle \hat{\tilde{x}}(q_{1},s_{1})\hat{\tilde{x}}(q_{2},s_{2})\rangle =
2\pi i \sum_{b=1}^{2} \mathop{\mathrm{Res}}_{z = z_b} F(z) \, .
\label{eq:S15}
\end{equation}
The residue of the integrand $F(z)$ at any pole $z_{b}$ is given by
\begin{equation}
\mathop{\mathrm{Res}}_{z = z_b} F(z) = \frac{\phi^{(m-1)}(z_b)}{(m-1)!} \, ,
\label{eq:16}
\end{equation}
where $m$ is the order of the pole and $\phi(z) = F(z_b)(z-z_b)^{m}$. Here, $\phi^{(m)}(z_b)$ is the $m^{th}$ derivative of $\phi$ with respect to $z$ at $z=z_b$.
Using contour integration, we evaluate the integral over $s_2$ to obtain,
\begin{equation}
\int_{c-i\infty}^{c+i\infty} ds_{2} e^{s_{2}t_{2}}\langle \hat{\tilde{x}}(q_{1},s_{1})\hat{\tilde{x}}(q_{2},s_{2})\rangle =
\frac{\tau_R\pi^2 i U_a^2\delta (q_2+q_1) }{(s_1+ q_1^2/\tau_H)(s_1 - q_2^2/\tau_H)}\big(e^{-q^2_2 t_2/\tau_H }
- e^{-s_1t_2}\big) \, ,
\label{eq:S17}
\end{equation}
and can rewrite Eq.~(\ref{eq:S13}) as
\begin{equation}
\langle \tilde{x}(q_{1}, t_{1})\tilde{x}(q_{2}, t_{2})\rangle = \frac{ \tau_R U_a^2\delta (q_2 +q_1)}{4i}\int_{c-i\infty}^{c+i\infty}ds_{1} \frac{e^{s_1t_1} e^{-q^2_2 t_2/\tau_H} - e^{s_1|t_1-t_2|}}{(s_1+ q_1^2/\tau_H)(s_1 - q^2_2/\tau_H)} \, .
\label{eq:S18}
\end{equation}
\begin{center}
\providecommand{\poles}{
  \node (poles) at (3.2,1.2) {poles of $e^{s_{2}t_{2}}\langle \hat{\tilde{x}}(q_{1},s_{1})\hat{\tilde{x}}(q_{2},s_{2})\rangle$};
  \draw[fill]
  (-.4, 0) coordinate [circle,fill,inner sep=1pt,label=below:$z_2$] (p1)
  (-.6, .5) coordinate [circle,fill,inner sep=1pt,label=below:$z_1$] (p2)
  (.5, 1.75) coordinate [circle,fill,inner sep=1pt,label=right:$c+i\infty$] (p3)
  (.5, -1.75) coordinate [circle,fill,inner sep=1pt,label=right:$c-i\infty$] (p4);
  \draw[ultra thin,gray] (poles) -- (p1) (poles) -- (p2);
}
\def\xr{1.2}
\def\yr{1.}
\begin{figure}
\centering
\begin{tikzpicture}[thick]
  \draw[->] (-\xr-1.,0) -- (\xr+1.,0) node [above right = -0.08] {$\text{Re}$};
  \draw[->] (0,-\yr-1.) -- (0,\yr+1.) node[above right= -0.08] {$\text{Im}$};
  \draw[fill] (0,0) circle (1pt) node [above right=0.1] {};
  \draw[fill] (-1.,1) circle (0pt) node [left] {$\Omega_{2}$};
  \draw[fill] (.6,-1) circle (0pt) node [right] {$\Omega_{1}$};
  \draw[xshift=15,blue!80!black,decoration={markings,mark=between positions 0.1 and 1 step 0.25 with \arrow{<}},postaction={decorate}] (0,\yr+0.75) -- (0,-\yr-0.75) node [above left] {$\Omega$} arc (270:90:\yr+0.75);
  \poles
\end{tikzpicture}
\caption{Integration contour in the complex plane corresponding to Eq.~(\ref{eq:S15})}
\label{Fig:1}
\end{figure}
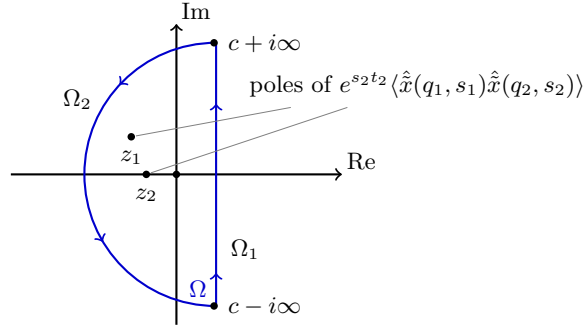
\end{center}
Here, we note that the system is time-translation invariant at steady-state, which is of our interest.
We include the absolute value of \( t_1 - t_2 \) to capture this fact and to ensure that the solution remains invariant under the order of integration over \( s_1 \) and \( s_2 \).
We also recognize that the integration in Eq.~(\ref{eq:S18}) can be evaluated using a similar procedure. 
Here, the integrand is redefined as
\[F(z) = \frac{e^{z t_1} e^{-q^2_2 t_2/\tau_H}- e^{z|t_1-t_2|}}{(z+q_1^2/\tau_H)(z - q^2_2/\tau_H)} \, , \]
and has two real simple poles at $z_{1} = -q^{2}_{1}/\tau_H$ and $ z_{2}=q^{2}_{2}/\tau_H$.
The integration over $\Omega_2$ vanishes due to Jordans's lemma for $t_1>0$ and $|t_1-t_2|>0$, and again integration over $\Omega_1$ is of primary interest. 
The integration over $\Omega_1$ is equal to the sum of the residues
\begin{equation}
\int_{\Omega_{1}} dz F(z)
= \int_{c-i\infty}^{c+i\infty} ds_{1} \frac{e^{s_1t_1} e^{-q^2_2 t_2/\tau_H}- e^{s_1|t_1-t_2|}}{(s_1+q_1^2/\tau_H)(s_1 - q^2_2/\tau_H)}
= 2\pi i \sum_{b=1}^{2} \mathop{\mathrm{Res}}_{z= z_b}F(z) \, ,
\label{eq:S19}
\end{equation}
and calculating the residues leads to the following expression,
\begin{equation}
\int_{c-i\infty}^{c+i\infty} ds_{1} \frac{e^{s_1t_1} e^{-q^2_2 t_2/\tau_H}- e^{s_1|t_1-t_2|}}{(s_1+ q_1^2/\tau_H)(s_1 - q^2_2/\tau_H)}
 = \frac{2\pi\tau_H i}{(q_1^2+q_2^2)} \bigg(-e^{-(q_1^2t_1+q_2^2t_2)/\tau_H} + e^{-q_1^2|t_1-t_2|/\tau_H} \bigg) \, .
\label{eq:S20}
\end{equation}
Using this result, we can write $\langle \tilde{x}(q_{1}, t_{1})\tilde{x}(q_{2}, t_{2})\rangle$ as
\begin{equation}
    \langle \tilde{x}(q_{1}, t_{1})\tilde{x}(q_{2}, t_{2})\rangle = \frac{\pi\tau_R\tau_H U_a^2 \delta (q_2+q_1)}{2(q_1^2+q_2^2)} \bigg(-e^{-(q_1^2t_1+q_2^2t_2)/\tau_H} + e^{-q_1^2|t_1-t_2|/\tau_H} \bigg) .
    \label{eq:S21}
\end{equation}
Upon taking the inverse Fourier transform of Eq.~(\ref{eq:S21}) and considering a single-particle correlation function ($n_1=n_2$), we obtain
\begin{equation}
    \langle x(t_{1})x(t_{2}) \rangle = \frac{\tau_R\tau_HU_a^2}{16\pi} \int_{-\infty}^{\infty}dq_{2} \frac{1}{q_2^2}\bigg(e^{- q_2^2|t_1-t_2|/\tau_H}- e^{-q_2^2(t_1+t_2)/\tau_H} \bigg) = \frac{1}{8}\tau_R U_a^2 \sqrt{\frac{\tau_H}{\pi}}\bigg( -\sqrt{|t_1-t_2|} + \sqrt{t_2+t_1} \bigg) \, .
    \label{eq:S22}
\end{equation}
We obtain our final result for the MSD at long-times by setting $t\equiv t_1 = t_2$ as
\begin{equation}
\langle\Delta x(t)^2\rangle = \langle x(t)x(t) \rangle = \frac{1}{8}\tau_R U_a^2 \sqrt{\frac{2\tau_H
t}{\pi}} \, ,
\label{eq:S23}
\end{equation}
which illustrates the subdiffusive behavior.


%